\documentstyle[12pt]{article}
\newtheorem{theorem}{Theorem}

\begin{document}

\author{ Marc Henneaux \\ %EndAName
Facult\'e des Sciences, Universit\'e Libre de Bruxelles,\\B-1050 Bruxelles,
Belgium\\and  \\ %EndAName
Centro de Estudios Cient\'\i ficos de Santiago,\\Casilla 16443, Santiago 9,
Chile}
\title{Homological Algebra and Yang-Mills Theory\thanks{%
Invited contribution to the 100th issue of Journal of Pure and Applied
Algebra marking the 25th Anniversary of the journal's existence and
dedicated to the applications of algebra to physics.}}
\date{December 12, 1994 }
\maketitle

\begin{abstract}
The antifield-BRST formalism and the various cohomologies associated with it
are surveyed and illustrated in the context of Yang-Mills gauge theory. In
particular, the central role played by the Koszul-Tate resolution and its
relation to the characteristic cohomology are stressed.
\end{abstract}

\newpage

\section{Introduction}

With the discovery by Becchi, Rouet and Stora \cite{BRS} and Tyutin \cite
{Tyutin} of the remarkable symmetry that now bears their names, it has been
realized that many fundamental questions of local field theory can be
reformulated as cohomological ones. This is true both at the classical and
at the quantum levels.

Reformulating physical problems as cohomological issues is more than just
aesthetically appealing since one can then analyse them by using the
powerful machinery of homological algebra. For instance, the homological
point of view has enabled us in \cite{H1} to streamline the demonstration of
a fundamental theorem of pertubative quantum Yang-Mills theory \cite
{Joglekar,Collins} and, more recently \cite{BH1,BBH1}, to definitely settle
a long-standing conjecture due to Kluberg-Stern and Zuber \cite{Kluberg,ZJ}
on the structure of renormalized gauge invariant operators.

There exists various formulations of the BRST symmetry. We shall follow in
this paper the approach that is now known as the ``antifield formalism''. A
detailed exposition of the antifield-BRST construction with an emphasis on
its cohomological aspects can be found in \cite{HT2}.

The distinguishing feature of the antifield formalism is the introduction of
new variables, the ``antifields''. These variables serve a definite purpose,
namely, they generate the ``Koszul-Tate'' differential complex, which
provides a (homological) resolution of the algebra of on-shell functions%
\footnote{%
By definition, the algebra of on-shell functions is the algebra of functions
defined on the stationary surface, i.e., on the surface where the equations
of motion hold.}. The Koszul-Tate complex plays a crucial role for at least
two reasons. First, its acyclicity properties guarantee the existence of the
BRST differential for an arbitrary gauge system, no matter how complicated
the structure of its gauge algebra is. Second, the Koszul-Tate differential
is a central tool in the calculation of the BRST cohomology.

The aim of this paper is to illustrate this second aspect of the Koszul-Tate
differential in the physically important case of the Yang-Mills theory. It
is sometimes stated that the antifield formalism is an unnecessary
complication for Yang-Mills models, whose gauge transformations are
irreducible and close off-shell. While this is certainly true for the
definition of the BRST symmetry itself, which can be given without ever
mentioning the antifield formalism, the full calculation of the BRST
cohomology is greatly simplified (and actually has been carried out only) by
following the homological ideas underlying the antifield theory. On these
grounds, the Yang-Mills models provide a nice illustration of the power of
the homological techniques applied to local field theory.

The antifield construction of the BRST differential finds its roots in the
papers \cite{Kallosh,deWit} and is due to \cite{BV,VT}. The introduction of
the antifields in the Yang-Mills case was actually performed earlier by
Zinn-Justin and Becchi, Rouet and Stora in \cite{ZJ2,BRS}, where the
antifields are called ``sources for the BRST variations of the fields''. The
key importance of the Koszul-Tate complex in the antifield formalism has
been uncovered and stressed in \cite{FH,H3}, where various cohomologies
occuring in BRST theory are also defined and related with one another. Our
approach to the antifield formalism follows the lines of our earlier work on
the Hamiltonian formulation of the BRST theory \cite{H2,HT1,FHST}, where
identical algebraic features hold (see \cite{HT2} for a unified and
systematic exposition).

\section{Yang-Mills Action}

As any local field theory with a gauge freedom, the Yang-Mills system is
characterized by its dynamics (partial differential equations deriving from
a variational principle) and by its symmetries. These two aspects are of
course connected since the symmetries leave, by definition, the action
invariant. The different cohomologies of interest to the analysis of a gauge
system are either associated with its dynamics (characteristic cohomology,
Koszul-Tate cohomology), to its symmetries (Lie algebra cohomology in
appropriate representation spaces) or to both (BRST cohomology).

The Yang-Mills action is given by
\begin{eqnarray}
  S[A_\mu ^a] & = & -\frac 14\int F_{\mu \nu }^aF^{b\mu \nu } \delta_{ab} d^nx
\label{action}
\end{eqnarray}
where $A_\mu ^a$ is the Yang-Mills connection. The field strengths read
explicitly
\begin{eqnarray}
   F^a_{\mu \nu}   & = & \partial _\mu A_\nu ^a-
\partial _\nu A_\mu ^a-C_{bc}^aA_\mu ^bA_\nu ^c.
   \label{strength}
\end{eqnarray}
Here, $C_{bc}^a$ are the structure constants of the Lie algebra ${\cal G}$
of the gauge group $G$. We shall assume that $G$ is compact and is thus the
direct product of a semi-simple compact Lie group $H$ by abelian factors,
\begin{eqnarray}
   G   & = & U(1)^k\times H.
   \label{group}
\end{eqnarray}
Couplings to matter will not be considered here for simplicity but can be
handled along the same lines \cite{BBH1}. The equations of motion following
from the action (\ref{action}) are the standard Yang-Mills equations,
\begin{eqnarray}
   D_\mu F^{a\mu \nu } & = & 0,
\label{YMequ}
\end{eqnarray}
where $D_\mu $ denotes the covariant derivative.

Let us now turn to the gauge symmetries of Yang-Mills theory. These are
given by
\begin{eqnarray}
  \delta _\epsilon A_\mu ^a   & =
 & \partial _\mu \epsilon^a -C_{bc}^aA_\mu ^a\epsilon ^c.
\label{gaugetran}
\end{eqnarray}
The invariance of the Yang-Mills action under the gauge transformations (\ref
{gaugetran}) is a direct consequence of the Noether identities
\begin{eqnarray}
   D_\mu \left( \frac{\delta {\cal L}}{\delta A_\mu ^a}\right)
\equiv  D_\mu D_\nu F^{a\mu \nu } \equiv 0.
\label{Noether}
\end{eqnarray}

\section{Algebra of Local Forms}

The differentials that we shall introduce are defined in the algebra of
local forms. A local form is by definition a form on spacetime (assumed to
be $R^n$ for simplicity), which depends also on the fields $A_\mu ^a$ and a
finite number of their derivatives. This can be formalized by using the jet
space language. We refer the reader to \cite{Anderson} for a presentation of
jet bundle theory adapted to our purposes and for further references. Let $E$
be the bundle $E=$ $R^n\times F$ over spacetime of Lie algebra-valued
exterior forms. Coordinates on the fibers are given by $A_\mu ^a$. A section
of $E$ is a field configuration $A_\mu ^a(x)$. The $k$-th order jet bundle
over $E$ is denoted by $J^k(E)$. It is coordinatized by $x^\mu $, the
components $A_\mu ^a$ of the vector potential and their successive
derivatives up to order $k$. The infinite jet bundle over $E$ is denoted by $%
J^\infty (E)$. The exterior algebra $\Omega (J^\infty (E))$ splits according
to vertical and horizontal degrees, $\Omega (J^\infty (E))=\oplus
_{r,s}\Omega ^{r,s}(J^\infty (E))$ where $r$ is the horizontal degree and $s$
is the vertical degree. The algebra of local forms is by definition the
algebra $\Omega ^{*,0}(J^\infty (E))$ of purely horizontal forms, i.e., of
elements of $\Omega (J^\infty (E))$ with zero vertical degree. A generic
element $\omega $ of $\Omega ^{*,0}(J^\infty (E))$ can be written as
\begin{eqnarray}
  \omega    & = & \sum_{0\leq p\leq n}\omega ^{(p)}
   \label{localforms}
\end{eqnarray}
with
\begin{eqnarray}
  \omega ^{(p)}  & = &  \sum \frac 1{p!}\omega _{\mu_1 \dots \mu_p}
dx^{\mu_1} \dots dx^{\mu_p}
\label{components}
\end{eqnarray}
(we do not write the exterior product $\wedge $ explicitly). The
coefficients $\omega _{\mu _1\dots \mu _p}$ are functions of the spacetime
coordinates, of the fields $A_\mu ^a$ and of a finite number of their
derivatives. We shall consider in the sequel only polynomial functions of
the $A_\mu ^a$ and their derivatives. Thus, the algebra ${\cal E}$ of local
forms can be identified with the tensor product
\begin{eqnarray}
  {\cal E}  & = & \Omega (R^n)\otimes  {\cal A},
\label{product}
\end{eqnarray}
where $\Omega (R^n)$ is the algebra of exterior forms on $R^n$ and where $%
{\cal A}$ is the algebra of ``local functions'', i.e., of (polynomial)
functions of the fields $A_\mu ^a$ and of a finite number of their
derivatives.

We shall introduce later further fields, which are the ghost fields and the
antifields. This means that the fibers $F$ of $E$ will be replaced by new
fibers denoted by $\bar F$ and coordinatized not only by $A_\mu ^a$ but also
by the new fields. The concept of local forms and local functions will be
modified accordingly, namely, the coefficients $\omega _{\mu_1\dots \mu_p} $
in (\ref{components}) will depend also (polynomially) on the ghosts, the
antifields and a finite number of their derivatives.

In the algebra of local forms, one can define the differential d as follows,
\begin{eqnarray}
   d\omega   & = & \sum_p d\omega ^{(p)} ,\\
   d\omega ^{(p)}   & = & \sum \frac 1{p!}
\left( d\omega _{\mu_1\dots \mu_p}\right) dx^{\mu_1} \dots dx^{\mu_p} ,\\
   d\omega _{\mu_1\dots \mu_p}   & = &dx^\mu \,
\partial _\mu \omega _{\mu_1\dots \mu_p} .
\end{eqnarray}
It coincides with the horizontal $d$ of the jet space formulation. To
describe the cohomology of $d$, it is convenient to introduce a further
concept. One says that a $n$-form $\omega \equiv adx^0\dots dx^{n-1}$ is
variationally closed if and only if the variational derivatives of a with
respect to all fields (and antifields if any) identically vanish,
\begin{eqnarray}
   \omega \equiv adx^{0} \dots dx^{n-1}
   \hbox{ is variationally closed}&
   \Longleftrightarrow  & \frac{\delta a}{\delta \Phi ^A}\equiv 0.
\label{varclosed}
\end{eqnarray}
Here, $\Phi ^A$ stands for the $A_\mu ^a$, as well as the ghosts and the
antifields if these are present. Furthermore, $\delta a/\delta \Phi ^A$ is
the variational derivative of $a$ with respect to $\Phi ^A$,
\begin{eqnarray}
   \frac{\delta a}{\delta \Phi ^A}  & \equiv &
\frac{\partial a}{\partial \Phi ^A}-
\partial _\mu \frac{\partial a}{\partial \left( \partial _\mu \Phi ^A\right)
}+\partial _\mu \partial _\nu \frac{\partial a}
{\partial \left( \partial _\mu \partial _\nu \Phi ^A\right) }- \dots
\label{varder}
\end{eqnarray}
Two $n$-forms are said to be equivalent if their difference is variationally
closed.

\begin{theorem}
(``Algebraic Poincar\'e Lemma''): The cohomology of $d$ is given by
\begin{eqnarray}
   H^0(d)   & = & R; \\
   H^k(d)   & = & 0; \\
   H^n(d)   & = & \hbox{ \{equivalent $n$-forms\}}.
\end{eqnarray}
\end{theorem}

\noindent
{\bf Proof} : This is a classical result from the calculus of variations,
see e.g. \cite{Anderson}.

\vspace{0.3cm}

We shall also need the cohomology of d in the algebra of invariant (local)
forms. A local form depending on $A_\mu ^a$ and its derivatives is gauge
invariant iff its coefficients involve only the field strength components
and their covariant derivatives contracted with tensors invariant for the
adjoint representation of the gauge group $G$ ( invariant polynomials in $%
F_{\mu \nu }^a$ and $D_\rho F_{\mu \nu }^a$, $D_\sigma D_\rho F_{\mu \nu }^a$%
, etc). The exterior derivative of an invariant local form is also an
invariant local form. Accordingly, one may investigate the invariant
cohomology of $d$, i.e., the cohomology of $d$ in the algebra of invariant
local forms. We shall call somewhat loosely ``characteristic classes'' the
invariant polynomials in the curvature two forms $F^a=\left( 1/2\right)
F_{\mu \nu }^adx^\mu dx^\upsilon $, e.g. $\delta _{ab}F^aF^b$ (where $\delta
_{ab}$ is the invariant Killing metric) is a characteristic class. The
characteristic classes are closed and thus exact by the algebraic Poincar\'e
lemma. However, they cannot be written as the $d$ of an invariant local
form. Indeed, one can prove~:

\begin{theorem}
(``Covariant Poincar\'e Lemma''): The invariant cohomology of $d$ in form
degree $k<n$ is given by the characteristic classes. In degree $n$, it
contains both the characteristic classes of degree $n$ (if any) and the
equivalence classes of invariant $n$-forms that differ by a variationally
closed one.
\end{theorem}

\noindent
{\bf Proof} : See \cite{BDK,DVHTV}.

\vspace{0.3cm}

There is an analogous theorem in Riemannian geometry, see \cite{Gilkey}. The
covariant Poincar\'e lemma provides a very nice algebraic characterization
of the characteristic classes.

\section{Characteristic Cohomology}

The next cohomology that we shall discuss is related to the equations of
motion. It is the characteristic cohomology \cite{Bryant}. To illustrate the
idea, we start with the familiar concept of conserved currents. A conserved
current $j$ may be defined as a $(n-1)$-form that is closed when the
equations of motion hold. We say that $j$ is closed ``on-shell'' and we
write
\begin{eqnarray}
  dj  & \approx & 0.
\label{cocyc1}
\end{eqnarray}
One easily constructs solutions of (\ref{cocyc1}) by taking simply
\begin{eqnarray}
    j & \approx  & dk,
\label{cobound1}
\end{eqnarray}
where $k$ is a $(n-2)$-form. These solutions are called trivial. The
characteristic cohomology in form degree $n-1$ is the quotient space of the
cocycles (\ref{cocyc1}) by the coboundaries (\ref{cobound1}).

More generally, the characteristic cohomology in form degree $k$ is defined
as the space of equivalence classes of cocycles $a$ solutions of
\begin{eqnarray}
   da  & \approx & 0
\label{cocyc2}
\end{eqnarray}
modulo the coboundaries $c$ defined by
\begin{eqnarray}
   c  & \approx & db.
\label{cobound2}
\end{eqnarray}
That is, the characteristic cohomology contains the non-trivial $k$-forms
that are closed on-shell.

If the gauge group G has abelian factors, the characteristic cohomology in
form degree $n-2$ does not vanish and contains the (equivalence classes) of
the cocycles $^{*}F^A$, where $A$ ranges over the abelian factors and where $%
^{*}F^A$ is the $n-2$-form dual of the abelian field strength $2$-form $F^A$%
. Indeed, the Maxwell equations for the abelian gauge fields read
\begin{eqnarray}
d (^*F^A) &\approx & 0,
\end{eqnarray}
and clearly $^{*}F^A$ is not equal on-shell to the exterior derivative of a
{\em local} form. We shall indicate below that there are no other non
trivial cocycles in form degree $n-2$. Furthermore, general arguments
establish the following theorem :

\begin{theorem}
(Vanishing Theorem for the Characteristic Cohomology in Degree $\leq n-3$):
The characteristic cohomology for Yang-Mills theory in form degree $k<n-2$
vanishes, except, of course, in degree $0$, where the characteristic
cohomology is isomorphic to the constants.
\end{theorem}

\noindent
{\bf Proof} : See \cite{Bryant,BBH2}.

\vspace{0.3cm}

Thus, the characteristic cohomology for Yang-Mills theory is completely
known up to form degree $n-2$ included. In form degree $n-1$, one has the
conserved currents. These probably reduce to the currents associated with
the Poincar\'e transformations (and in 4 dimensions, with the conformal
symmetries as well) \cite{Torre}. Restrictions on the form of the conserved
currents are given in \cite{BBH3}.

\section{Koszul-Tate Complex}

The second differential associated with the equations of motion is the
Koszul-Tate differential $\delta $. The equations of motion (\ref{YMequ}),
together with their successive derivatives up to order $k-2$,
\begin{eqnarray}
   D_\nu F^{a \mu \nu}  & = & 0 \\
   \partial_{\lambda_1 \dots \lambda_i} D_\nu F^{a \mu \nu}   & = & 0,
\; i = 1, \dots , k-2
\label{prolonged}
\end{eqnarray}
determine a surface $\Sigma _k$ in $J^k(E)$ ($k=2,3,\dots $). They form what
is known as the ``prolonged system'' and are subject to the Noether
identities (\ref{Noether}) together with their derivatives up to order $k-3$%
,
\begin{eqnarray}
   D_\mu D_\nu F^{a\mu \nu }  & = & 0 \\
  \partial _\lambda D_\mu D_\nu F^{a\mu \nu } & = & 0, \hbox{ etc}.
   \label{Noetherbis}
\end{eqnarray}
The Koszul-Tate differential implements the equations of motion in
cohomology. Technically, this means that it provides for each value of $k$ a
resolution of the algebra $C^\infty (\Sigma _k)$ of functions on $\Sigma _k$%
. It is defined on the generators of $C^\infty (J^k(E))\otimes \Lambda
\left[ A_a^{*\mu },\partial _\lambda A_a^{*\mu },\dots ,\partial _{\lambda
_1 \lambda _2 \dots  \lambda _{k-2}}A_a^{*\mu }\right]
\otimes R\left[ C_a^{*},\partial _\lambda C_a^{*},\partial _{\lambda
_1 \lambda _2 \dots \lambda _{k-3}}C_a^{*}\right] $ as
follows
\begin{eqnarray}
  \delta f & = & 0 \; \; \forall f\in C^\infty (J^k(E)),\\
   \delta (\partial _{\lambda _1}
\partial _{\lambda _2}\dots \partial _{\lambda _i}A_a^{*\mu })  & = &
 \partial _{\lambda _1} \partial _{\lambda _2}\dots \partial _{\lambda _i}
D_\nu F_a^{\mu \nu} \\
   \delta (\partial _{\lambda _1}\partial _{\lambda _2}\dots
   \partial _{\lambda _j}C_a^{*})
& = & \partial _{\lambda _1}\partial _{\lambda _2}
\dots \partial _{\lambda _j} \, D_\mu A_a^{*\mu }
\end{eqnarray}
and extended to the full algebra as an odd derivation. The $A_a^{*\mu }$ and
$C_a^{*}$ are the ``antifields''; they are respectively odd and even and
have the following gradings
\begin{eqnarray}
  antigh(A^{*\mu}_a)   & = &1 \\
  antigh(C_a^*)   & = &2
\end{eqnarray}
(of course $antigh(A_\mu ^a)=0$). One has $\delta ^2=0$ because of the
Noether identities (\ref{Noether}). Thus, $\delta $ is a differential of
antighost number $-1$. Furthermore, $\delta $ commutes with $\partial _\mu $%
, which implies $\delta d+d\delta =0$.

The cohomology of the Koszul-Tate differential $\delta $ is given by the
following theorem.

\begin{theorem}
(Cohomology of Koszul-Tate differential $\delta $): For each $k$, the
Koszul-Tate complex $\left( K_k,\delta \right) $ of order $k$, with
\begin{eqnarray}
   K_k \equiv C^\infty (J^k(E))\otimes \Lambda \left[ A_a^{*\mu },
\dots ,\partial _{\lambda _1 \lambda _2 \dots \lambda _{k-2}}A_a^{*\mu }\right]
\otimes R\left[ C_a^{*},\dots,
\partial _{\lambda _1 \lambda _2 \dots
\lambda _{k-3}}C_a^{*}\right]
\label{KTate}
\end{eqnarray}
provides a resolution of the algebra $C^\infty (\Sigma _k)$ of functions
defined on the stationary surface. That is,
\begin{eqnarray}
   H_0(\delta)   & = &C^\infty (\Sigma _k), \label{resolu} \\
   H_j(\delta)   & = & 0,\; \; j>0.
   \label{acydelta}
\end{eqnarray}.
\end{theorem}

\noindent
{\bf Proof} : See \cite{FHST,FH,H3,H4}.

\vspace{0.3cm}

It follows from this theorem that the equation $f\approx 0$, where $f$ has
antighost number $0$, is equivalent to $f=\delta m$ for some $m$ of
antighost number $1$. One can replace $C^\infty (J^k(E))$ in $K_k$ by $%
\Omega ^{*,0}(J^\infty (E))$, i.e., one can take the tensor product of $K_k$
with $\Lambda \left[ dx^\mu \right] $. The theorem remains then true
provided one replaces in (\ref{resolu}) the algebra $C^\infty (\Sigma _k)$
by $C^\infty (\Sigma _k)\otimes \Lambda \left[ dx^\mu \right] $. Also, one
may consider -as we do here - polynomial functions on the jet spaces without
changing the conclusions since the equations of motion themselves are
polynomial.

Another cohomology of interest is the cohomology $H_i^j(\delta |d)$ of $%
\delta $ modulo $d$. Here, $j$ is the form degree while $i$ is the antighost
number. The cohomology $H_{*}^{*}(\delta |d)$ is defined by the cocycle
condition
\begin{eqnarray}
  \delta a + db   & = &0
  \label{cocydeltad}
\end{eqnarray}
with the coboundary condition
\begin{eqnarray}
  a \hbox{ is a coboundary for $\delta$ modulo $d$ }  &\Leftrightarrow  &a =
\delta c + d e.
\label{coboundeltad}
\end{eqnarray}
This cohomology arises when studying the cohomology of $\delta $ in the
space of local functionals, i.e., in the space of integrated local $n$-forms
$\int \omega ^{\left( n\right) }$ \cite{H4}. Contrary to $H(\delta )$, it
may not vanish for antighost number $i\neq 0$ \cite{H4}. In fact, in the
Yang-Mills case, one has :

\begin{theorem}
(Cohomology of $\delta $ modulo $d$, $n>2$): The only non-vanishing
cohomological groups $H_i^j(\delta |d)$, $i>0$, $j>0$ are $H_2^n(\delta |d)$%
, $H_1^n(\delta |d)$ and $H_1^{n-1}(\delta |d)$. The group $H_1^n(\delta |d)$
is isomorphic to the set of equivalence classes of symmetries, where two
symmetries (i.e., two transformations of the $A_\mu ^a$ leaving the
Yang-Mills action invariant) are identified if their difference reduces to a
gauge transformation on-shell. Furthermore, $H_2^n(\delta |d)$ and $%
H_1^{n-1}(\delta |d)$ are isomorphic. A complete list of independent
representatives of the classes of $H_2^n(\delta |d)$ is given by
\begin{eqnarray}
  C^*_A dx^0 dx^1 \dots dx^{n-1}
\label{H2}
\end{eqnarray} where $A$ ranges over the abelian factors of the gauge group.
\end{theorem}

\noindent
{\bf Proof} : See \cite{BBH2}.

\vspace{0.3cm}

The isomorphism of $H_2^n(\delta |d)$ and $H_1^{n-1}(\delta |d)$ follows
from standard descent equation techniques and from the vanishing theorems
for $H(\delta )$ and $H(d)$. For a semi-simple gauge group, $H_2^n(\delta
|d) $ and $H_1^{n-1}(\delta |d)$ vanish. Only $H_1^n(\delta |d)$ is non
trivial.

\section{A New Look at the Characteristic Cohomology}

The characteristic cohomology can be reformulated as a ``mod'' cohomology
with the help of the Koszul-Tate differential. Indeed, the cocycle condition
(\ref{cocyc2}) can be rewritten as
\begin{eqnarray}
  da+\delta m  & = & 0
\label{cocyc5}
\end{eqnarray} for some form $m$ of antighost number $1$. Similarly, the
coboundary condition (\ref{cobound2}) is equivalent to
\begin{eqnarray}
  c  & = & db +\delta n.
\label{cobound5}
\end{eqnarray}for some $n$ of antighost number $1$. Thus, the characteristic
cohomology is just the cohomology $H_0^j(d|\delta )$ of $d$ modulo $\delta$.
We shall in the sequel refer to $H_i^j(d|\delta )$ as the characteristic
cohomology even when $i\neq 0$.

The vanishing theorems for $H(\delta )$ and $H(d)$ easily lead to the
following isomorphism between the characteristic cohomology and $%
H_i^j(\delta |d)$.

\begin{theorem}
(Isomorphism Between the Characteristic Cohomology and \\
$H_i^j(\delta |d)$):
One has
\begin{eqnarray}
   H^i_j(\delta | d)   & \simeq  & H^{i-1}_{j-1}(d|\delta) \;
(i>0,\, j>0,\, (i,j) \not= (1,1)); \\
   H^1_1(\delta | d)   & \simeq & \frac{H^0_0(d | \delta)}{R}.
\end{eqnarray}
\end{theorem}

\noindent
{\bf Proof} : See \cite{BBH2}.

\vspace{0.3cm}

In particular, the isomorphism $H_0^{n-1}\left( d|\delta \right) \simeq
H_1^n\left( \delta |d\right) $ is just a cohomological reformulation of
Noether theorem since $H_0^{n-1}\left( d|\delta \right)$ contains the
conserved currents and $H_1^n\left( \delta |d\right) $ the symmetries of the
action. Moreover, the elements in $H_2^n\left( \delta |d\right) $
corresponding to $^{*}F_A\in H_0^{n-2}\left( d|\delta \right) $ through the
isomorphisms $H_0^{n-2}\left( d|\delta \right) \simeq H_1^{n-1}\left( \delta
|d\right) \simeq H_2^n\left( \delta |d\right) $ are just the $n$-forms $%
C_A^{*}dx^0...dx^{n-1}$ of equation (\ref{H2}). It is actually by using the
isomorphism theorem and computing directly $H_2^n\left( \delta |d\right) $
that we have been able to work out the characteristic cohomology in degree $%
n-2$ for Yang-Mills theory in reference \cite{BBH2}.

\section{Lie Algebra Cohomology}

We now introduce a differential which, contrary to the previous ones, is
related to the gauge transformations rather than to the equations of motion.

The gauge transformations (\ref{gaugetran}) act on the bundle $E\equiv
J^0(E) $. They provide a non-linear realization of the Lie algebra ${\cal G}%
_1$ parametrized by $\epsilon ^a$ and $\partial _\mu \epsilon ^a$, regarded
as independent parameters. The commutator of two elements $(\epsilon
^a,\partial _\mu \epsilon ^a)$ and $(\eta ^a,\partial _\mu \eta ^a)$ is
parametrized by $(\zeta ^a,\partial _\mu \zeta ^a)$, with
\begin{eqnarray}
   \zeta ^a   & = & C_{bc}^a\epsilon ^b\eta ^c.
\label{comm}
\end{eqnarray}
The Lie algebra ${\cal G}\equiv {\cal G}_0$ of the gauge group is the
subalgebra with $\partial _\mu \epsilon ^a=0$ (constant gauge
transformations). The action is non linear because $\delta _\epsilon A_\mu
^a $ does not depend linearly on $A_\mu ^a$ in (\ref{gaugetran}): if one
multiplies $A_\mu ^a$ by $5$, say, $\delta _\epsilon A_\mu ^a$ is not
multiplied by $5$ because of the inhomogeneous term $\partial _\mu \epsilon
^a$.

One can extend the action of the gauge transformations (\ref{gaugetran}) to
any finite order jet bundle $J^k(E)$ by simply taking their successive
derivatives with respect to the coordinates $k$ times (``prolongation of the
symmetry" in jet bundle terminology). For instance, in $J^1(E)$, the gauge
transformations are given by (\ref{gaugetran}) and
\begin{eqnarray}
  \delta _\epsilon \left( \partial _\mu A_\nu ^a\right) \equiv
\partial _\mu \delta _\epsilon A_\nu ^a  & = & \partial _{\mu \nu }\epsilon ^a-
C_{bc}^a\partial _\mu A_\nu ^b\epsilon ^c-
C_{bc}^aA_\nu ^b\partial _\mu \epsilon ^c.
\label{gaugetran2}
\end{eqnarray}
They are parametrized by $(\epsilon ^a,\partial _\mu \epsilon ^a,\partial
_\nu \partial _\mu \epsilon ^a)$ and form a Lie algebra ${\cal G}_2$ of
which ${\cal G}_0$ is again the subalgebra with gauge parameters having
vanishing derivatives. The commutator of two elements $(\epsilon ^a,\partial
_\mu \epsilon ^a,\partial _\nu \partial _\mu \epsilon ^a)$ and $(\eta
^a,\partial _\mu \eta ^a,\partial _\nu \partial _\mu \eta ^a) $ of ${\cal G}%
_2$ is given by $(\zeta ^a,\partial _\mu \zeta ^a,\partial _\nu \partial
_\mu \zeta ^a)$ with $\zeta ^a$ given again by (\ref{comm}). We shall denote
by ${\cal G}_{k+1}$ the Lie algebra acting in $J^k(E)$. One can even go all
the way to the infinite jet space $J^\infty (E)$. The corresponding Lie
algebra of gauge transformations is parametrized by $\epsilon ^a$ and all
its derivatives, and is denoted by ${\cal G}_\infty $.

The Lie algebra ${\cal G}_0$ of the gauge group is more than just a
subalgebra of ${\cal G}_\infty $. It is also the quotient algebra of ${\cal G%
}_\infty $ by the infinite dimensional ideal ${\cal G}_\infty ^{^{\prime }}$
of transformations with $\epsilon ^a=0$ (but $\partial _\mu \epsilon ^a\neq
0 $, $\partial _\nu \partial _\mu \epsilon ^a\neq 0$, etc)
\begin{eqnarray}
  {\cal G}_0   & = & \frac {{\cal G}_\infty }{{\cal G}_\infty ^{^{\prime }}}.
\end{eqnarray}
One calls tensor representations of ${\cal G}_\infty $ the linear
representations of ${\cal G}_\infty $ in which the subalgebra ${\cal G}%
_\infty ^{^{\prime }}$ is mapped on zero. These are in bijective
correspondence with the linear representations of ${\cal G}_0$. The
curvature components $F_{\mu \nu }^a$ and their successive covariant
derivatives, for fixed values of the spacetime indices, transform all in the
tensor representation of ${\cal G}_\infty $ corresponding to the adjoint
representation of ${\cal G}_0$.

Because ${\cal G}_\infty $ acts on $J^\infty (E)$ and thus also on functions
on $J^\infty (E)$, one can introduce in the usual manner the coboundary
operator $\gamma $ for the Lie algebra cohomology in the module of functions
on $J^\infty (E)$ through the formula
\begin{eqnarray}
  \gamma \partial _{\mu _1\mu _2\dots \mu _k} A^a_\mu  & = &
\partial _{\mu _1\mu _2...\mu _k} D_\mu C^a,\\
\gamma \partial _{\mu _1\mu _2\dots \mu _k}C^a & = &
\partial _{\mu _1\mu _2\dots \mu _k}\left( \frac 12 C_{bc}^a C^b C^c\right) .
\end{eqnarray}
One extends $\gamma $ to the algebra $C^\infty (J^\infty (E))\otimes \Lambda
\left[ C^a,\partial _\mu C^a,\partial _{\mu \nu }C^a...\right] $ as an odd
derivation. The odd generators $C^a$ are known as the ghosts and are
assigned pure ghost number equal to $1$, whereas $A_\mu ^a$ has pure ghost
number equal to zero. Accordingly, $\gamma $ has pure ghost number equal to $%
1$. It is immediate to verify that $\gamma \partial _\mu =\partial _\mu
\gamma $. Since the action of $\gamma $, although non linear, is
nevertheless polynomial, one may restrict $C^\infty (J^\infty (E))$ to
polynomial functions.

One may then extend the action of $\gamma $ to the antifields by demanding  $%
\gamma \delta +\delta \gamma =0$ and $\delta \partial _{\mu _1\mu _2\dots
\mu _k}C^a=0$. Because the left-hand sides $D_\nu F_a^{\mu \nu }$ of the
Yang-Mills equations of motion transform tensorially in the coadjoint
reprentation of ${\cal G}_0$, this requirement is equivalent to stating that
the antifields (and their covariant derivatives) transform also in the
tensorial representation of ${\cal G}_\infty $ corresponding to the
coadjoint representation of ${\cal G}_0$. Consequently,
\begin{eqnarray}
   \gamma  A_a^{*\mu }  & = &  C_{ab}^c  C^b A_c^{*\mu}, \\
  \gamma  C_a^{*}  & = & C_{ab}^c  C^b C_c^*.
\end{eqnarray}
The action of $\gamma $ on the derivatives of the antifields is obtained by
the rule $\gamma \partial _\mu =\partial _\mu \gamma $. To complete the
definition of the differential $\gamma $ in the (polynomial) algebra $\bar {%
{\cal E}}$ of local forms involving the vector potentials, the antifields,
the ghosts and their derivatives, one sets $\gamma dx^\mu =0$, which implies
$\gamma d+d\gamma =0$. The antifields are of course assigned pure ghost
number equal to zero, while the ghosts have antighost number zero (see e.g.
\cite{BBH1} for more information).

The cohomology of $\gamma $ in the algebra $\bar {{\cal E}}$ is the Lie
algebra cohomology of the infinite dimensional Lie algebra ${\cal G}_\infty $
in the module of the polynomials in the fields $A_\mu ^a$, the antifields
and their derivatives. To compute it more explicitly, it is necessary to
understand the role played by the non linear term $\partial _\mu C^a$
appearing in the transformation law of $A_\mu ^a$. To that end, let us first
consider the case of a single abelian field $A_\mu $. One then has
\begin{eqnarray}
  \gamma \partial _{\mu _1\mu _2 \dots \mu _k}A_\mu   & = &
\partial _{\mu \mu _1\mu _2\dots \mu _k}C,
\end{eqnarray}
$\delta $(everything else) $=0$. One sees that the symmetrized derivatives
of $A_\mu $ and the derivatives of the ghost form contractible pairs and
drop from the cohomology. Thus, the cohomology of the infinite-dimensional
Lie algebra ${\cal G}_\infty $ acting on the ${\cal G}_\infty $-module of
the polynomials in the $A_\mu $'s, the antifields and their derivatives
actually reduces to the cohomology of the finite-dimensional Lie algebra $%
{\cal G}_0$ acting on the ${\cal G}_0$-module of the polynomials in the $%
F_{\mu \nu }$'s, the antifields and their derivatives (which actually
transform all according to the trivial representation of ${\cal G}_0$).

The same property is true in the general case : the derivatives of the
ghosts are killed in cohomology by the symmetrized derivatives of the vector
potential. Let ${\cal C}$ be the algebra of local polynomial forms that
involve only the field strength components $F_{\mu \nu }^a$, the antifields $%
A_a^{*\mu }$, the antifields $C_a^{*}$ and their succesive covariant
derivatives $D_{\lambda _1 \dots \lambda _i}F_{\mu \nu }^a$, $D_{\lambda _1
\dots \lambda _i}A_a^{*\mu }$, $D_{\lambda _1 \dots \lambda _i}C_a^{*}$ ($%
i=1,2,3,\dots)$. We shall denote all these variables, which transform
tensorially, by $z^\Delta $. The algebra ${\cal C}$ is clearly the
representation space of a {\em tensor} representation of ${\cal G}_\infty $
. Thus, contrary to $\bar {{\cal E}}$, it provides also a representation of
the finite dimensional Lie algebra ${\cal G}_0$. The theorem that
generalizes the situation found in the abelian case is

\begin{theorem}
(Isomorphism Between $H(\gamma ,\bar {{\cal E}})$ and the Lie algebra
cohomology $H({\cal G}_0,{\cal C})$): The cohomology of the differential $%
\gamma $ in $\bar {{\cal E}}$, which is identical to the Lie algebra
cohomology $H({\cal G}_\infty ,\bar {{\cal E}})$ of the infinite-dimensional
Lie algebra ${\cal G}_\infty $ in the ${\cal G}_\infty $-module $\bar {{\cal %
E}}$, is isomorphic to the Lie algebra cohomology $H({\cal G}_0,{\cal C})$
of the finite-dimensional Lie algebra ${\cal G}_0$ in the ${\cal G}_0$%
-module ${\cal C}$,
\begin{eqnarray}
  H(\gamma ,\bar {\cal E})
  \equiv H({\cal G}_\infty ,\bar {\cal E}) & \simeq  &
H({\cal G}_0,{\cal C}).
\end{eqnarray}
\end{theorem}

\noindent
{\bf Proof} : See \cite{DVHTV,BDK,H1}.

\vspace{0.3cm}

Since the algebra ${\cal G}_0$ is finite-dimensioal, one can use standard
theorems on Lie algebra cohomology to find

\begin{theorem}
(Cohomology of $\gamma $): Up to $\gamma $-exact terms, the general solution
of the cocycle condition $\gamma a=0$ in ${\cal E}$ reads
\begin{eqnarray}
   a  & = & \sum P_J(z^\Delta )\omega ^J(C^a),
\end{eqnarray}
where the sum is over a basis $\left\{ \omega ^J\right\} $ of the Lie
algebra cohomology $H({\cal G}_0)\equiv H({\cal G}_0,{\rm R})$ of ${\cal G}_0
$ and where the $P_J$ are invariant polynomials in the $z^\Delta $ (which
may also involve the coordinates $x^\mu $ and the $dx^\mu $).
\end{theorem}

In the case of an abelian gauge group, there is just one $\omega ^J,$
namely, the abelian ghost $C$ itself. For $SU(m)$, the algebra $H({\cal G}%
_0) $ is generated by the ``primitive forms'' $trC^3$, $trC^5$ up to $%
trC^{2m-1}$. Here we have set $C\equiv C^aT_a$, where the $T_a$ are the
generators of the adjoint representation.\ \ Finally, we point out that the
cohomology $H(\gamma |d)$ of $\gamma $ modulo $d$ has also been investigated
in the literature \cite{BDK,DVHTV,BBH1}, but we shall not report the results
here.

\section{BRST Cohomology}

The BRST differential takes into account both the dynamics and the gauge
symmetries. For a theory with a gauge algebra that closes off-shell, it is
simply the sum of $\delta $ and $\gamma $,
\begin{eqnarray}
  s  & = & \delta + \gamma.
\end{eqnarray}
It increases by one unit the (total) ghost number, defined to be the
difference between the pure ghost number and the antighost number. The BRST
cohomology can be shown on general grounds to be equal to
\begin{eqnarray}
  H^k(s)  & \simeq  & H^k(\gamma ,H_0(\delta )),
\end{eqnarray}
where $k$ stands in the left-hand side for the total ghost number while it
stands in the right hand side for the pure ghost number. This follows from a
simple spectral sequence argument explained for example in this particular
context in \cite{HT2}.

In practice, however, one needs a more precise characterization of the
representatives of the BRST cohomology. This is given by the following
theorem.

\begin{theorem}
(``Joglekar and Lee Theorem''): In each equivalence class of the BRST
cohomology, one can find a representative that does not depend on the
antifields and that is thus annihilated by $\gamma $. In particular, if $a$
is a BRST cocycle of ghost number zero, then one has
\begin{eqnarray}
  sa = 0, \; gh(a) = 0 & \Leftrightarrow   &
a = \bar a + sb,
\end{eqnarray}
where $\bar a$ is an invariant polynomial in the field strength components
and their covariant derivatives.
\end{theorem}

\noindent
{\bf Proof} : See \cite{Joglekar,H1}.

\vspace{0.3cm}

This theorem plays a key role in the analysis of the renormalization of
local gauge invariant operators \cite{Collins}.

\section{Wess-Zumino Consistency Condition}

Another cohomology that is also of fundamental physical importance is the
cohomology $H(s|d)$ of $s$ modulo $d$. The corresponding cocycle condition
reads
\begin{eqnarray}
   sa + db    & = & 0
\label{WZcons}
\end{eqnarray}
and is known as the Wess-Zumino consistency condition \cite{WZ}. Trivial
solutions of (\ref{WZcons}) are of the form $a=sm+dn$. The Wess-Zumino
consistency condition arose first in the context of anomalies, where it
constraints severely the form of the candidate anomalies, but it is also
quite important in analyzing the counterterms that are needed in the
renormalization of integrated gauge invariant operators and, classically, in
determining the form of the consistent deformations of the action.

The general solution of the Wess-Zumino consistency condition for Yang-Mills
theory has been worked out in \cite{BBH1} and involves all the cohomologies
described previously. There is indeed a close connection between $H(s|d)$, $%
H(\delta |d)$ and $H(\gamma |d),$ given again by a spectral sequence
argument \cite{HT2,BBH2}. We shall not give here the results of our general
investigation but shall merely quote two theorems of direct physical
interest. These theorems are valid in {\em four spacetime dimensions} and
for a {\em semi-simple} gauge group (no abelian factors).

\begin{theorem}
(On the Kluberg-Stern and Zuber conjecture): Up to trivial terms, the
general solution of the Wess-Zumino consistency condition with ghost number
zero is exhausted by the invariant polynomials in the field strength
components and their covariant derivatives.
\end{theorem}

\noindent
{\bf Proof} : See \cite{BH1,BBH1}.

\vspace{0.3cm}

\begin{theorem}
(``Adler-Bardeen anomaly''): The only solutions of the Wess-Zumino
consistency condition with ghost number one are given by the multiples of
the Adler-Bardeen anomaly
\begin{eqnarray}
   a  & = & tr\left\{ C\left[ dAdA+
\frac 12\left( AdAA-A^2dA-dAA^2\right) \right] \right\}
\label{AdlerBardeen}
\end{eqnarray} (up to trivial terms). Here, $A$ is the one-form $A_\mu
^adx^\mu T_a$ .
\end{theorem}

\noindent
{\bf Proof} : See \cite{BH1,BBH1}.

\vspace{0.3cm}

These theorems are not true anymore in other spacetime dimensions (in odd
dimensions, one has the Chern-Simons forms that solve (\ref{WZcons}) at
ghost number zero), or in the presence of abelian factors. We refer to \cite
{BBH1} for more information.

\section{Conclusions}

In this paper, we have illustrated the usefulness and power of cohomological
concepts in local field theory. Even though its gauge structure is simple
(off-shell closure, irreducibility), the Yang-Mills system provides a great
diversity of cohomologies of direct physical significance. These
cohomologies are the basic ingredients of the antifield formalism and allow
one to provide the general solution of the crucial Wess-Zumino consistency
condition with antifields included. The approach to Yang-Mills gauge models
outlined here can be characterized as the ``cohomological approach to
Yang-Mills theory".

Following analogous ideas, a similar analysis of the Wess-Zumino consistency
condition has been carried out successfully for other theories with a gauge
freedom, including Einstein gravity \cite{BBH4}, two-dimensional gravity
\cite{BTVP} as well as theories involving  p-form gauge fields \cite{HKS}.

\section{ Acknowledgements}

The author is grateful to Glenn Barnich, Friedemann Brandt, Michel
Dubois-Violette, Jean Fisch, Jim Stasheff, Michel Talon, Claudio Teitelboim
and Claude Viallet for very useful discussions and collaborations. This work
has been supported in part by research funds from F.N.R.S. and by research
contracts with the Commission of the European Communities.


\newpage

\begin{thebibliography}{99}
\bibitem{BRS}  C. Becchi, A. Rouet and R. Stora, {\em Commun. Math. Phys.}
{\bf 42} (1975) 127; {\em Ann. Phys. (N.Y.)} {\bf 98} (1976) 287.

\bibitem{Tyutin}  I.V. Tyutin, Lebedev preprint FIAN {\bf 39} (1975).

\bibitem{H1}  M. Henneaux, {\em Phys. Lett.} {\bf B313} (1993) 35.

\bibitem{Joglekar}  S.D. Joglekar and B.W. Lee, Ann. Phys (N.Y.) 97 (1976)
160.

\bibitem{Collins}  J.C. Collins, {\em Renormalization}, Cambridge University
Press (Cambridge~: 1984); J.C. Collins and R.J. Scalise, ${\em Phys.Rev.}$\
{\bf D50} (1994) 4117; B.W. Harris and J. Smith, {\em Anomalous Dimension of
the Gluon Operator in Pure Yang-Mills Theory}, Stony Brook preprint
ITP-SB-94-39.

\bibitem{BH1}  G. Barnich and M. Henneaux, {\em Phys. Rev. Lett.} {\bf 72 }%
(1994) 1588.

\bibitem{BBH1}  G. Barnich, F. Brandt and M. Henneaux, hep-th/9405194, to
appear in {\em Commun. Math. Phys.}

\bibitem{Kluberg}  H. Kluberg-Stern and J.B. Zuber, {\em Phys. Rev.} {\bf D12%
} (1975) 467; {\bf D12} (1975) 482; {\bf D12} (1975) 3159.

\bibitem{ZJ}  J. Zinn-Justin, {\em Quantum Field Theory and Critical
Phenomena}, Clarendon Press (Oxford : 1993).

\bibitem{HT2}  M. Henneaux and C. Teitelboim, {\em Quantization of Gauge
Systems}, Princeton University Press (1992).

\bibitem{Kallosh}  R. E. Kallosh, {\em Nucl. Phys.} {\bf B141} (1978) 141.

\bibitem{deWit}  B. de Wit and J.-W. van Holten, {\em Phys. Lett.} {\bf 79B}
(1978) 389.

\bibitem{BV}  I.A. Batalin and G.A. Vilkovisky, {\em Phys. Lett.} {\bf B};
{\em Phys. Rev.} {\bf D28} (1983) 2567; {\em J. Math. Phys.} {\bf 26} (1985)
172.

\bibitem{VT}  B.L. Voronov and I.V. Tyutin, {\em Theor. Math. Phys.} {\bf 50
}(1982) 218.

\bibitem{ZJ2}  J. Zinn-Justin, ``Renormalization of Gauge Theories'', in
{\em Trends in Elementary Particle Theory, Lecture Notes in Physics}, n\o
{\bf 37}, ed. H. Rollnik and K. Dietz, Springer Verlag (Berlin : 1975).

\bibitem{FH}  J.M.L. Fisch and M. Henneaux, {\em Commun. Math. Phys.} {\bf %
128} (1990) 627.

\bibitem{H3}  M. Henneaux, {\em Nucl. Phys. B (Proc. Suppl.)} 18A (1990) 47.

\bibitem{H2}  M. Henneaux, {\em Phys. Rep.} {\bf 126} (1985) 1.

\bibitem{HT1}  M. Henneaux and C. Teitelboim, {\em Commun. Math. Phys.} {\bf %
115} (1988) 213.

\bibitem{FHST}  J. Fisch, M. Henneaux, J.D. Stasheff and C. Teitelboim, {\em %
Commun. Math. Phys.}{\bf 120} (1989) 379.

\bibitem{Anderson}  I.M. Anderson, {\em Introduction to the variational
bicomplex}, in {\em Mathematical Aspects of Classical Field Theory}, M.J.
Gotay, J.E. Marsden and V. Moncrief eds, Contemporary Mathematics (American
Mathematical Society : Rhode Island 1992).

\bibitem{BDK}  F. Brandt, N. Dragon and M. Kreutzer, {\em Nucl. Phys.} {\bf %
B332} (1990) 224.

\bibitem{DVHTV}  M. Dubois-Violette, M. Henneaux, M. Talon and C.M. Viallet,
{\em Phys. Lett.} {\bf B267} (1991) 81.

\bibitem{Gilkey}  P. Gilkey, {\em Adv. in Math.} {\bf 28} (1978) 1.

\bibitem{Bryant}  R.L. Bryant and P.A. Griffiths, {\em Characteristic
Cohomology of Differential Systems (I): General Theory}, Duke University
Mathematics Preprint series, volume 1993 n\o 1 (January 1993).

\bibitem{BBH2}  G. Barnich, F. Brandt and M. Henneaux, hep-th/9405109, to
appear in {\em Commun.\ Math. Phys}.

\bibitem{Torre}  C.G. Torre, hep-th/9407129.

\bibitem{BBH3}  G. Barnich, F. Brandt and M. Henneaux, hep-th/9411202.

\bibitem{H4}  M. Henneaux, {\em Commun. Math. Phys.} {\bf 140} (1991) 1.

\bibitem{WZ}  Wess and Zumino, {\em Phys.\ Lett.}\ {\bf 37B} (1971) 95.

\bibitem{BBH4}  G. Barnich, F. Brandt and M. Henneaux, hep-th/9409104 (to
appear in {\em Phys. Rev. D} as a Rapid Communication); longer version in
preparation.

\bibitem{BTVP}  F. Brandt, W. Troost and A. Van Proeyen, hep-th/94070621

\bibitem{HKS}  M. Henneaux, B. Knaepen and C. Schomblond, in preparation.
\end{thebibliography}
\end{document}